\documentclass[twocolumn,showpacs,preprintnumbers]{revtex4}
\usepackage{amssymb}
\usepackage{amsmath}
\usepackage{graphicx}
\usepackage{dcolumn}
\usepackage{bm}
\usepackage{subfigure}
\usepackage[colorlinks,citecolor=blue, linkcolor=blue,hyperindex,dvipdfm]{hyperref}

\begin{document}

\title{Thermodynamic properties of higher-dimensional dS black holes in dRGT
massive gravity}
\author{Yubo Ma$^{1}$, Yang Zhang$^{1}$, Lichun Zhang$^{1}$,  Liang Wu$^{1}$, Yumei Huang$^{2}$%
, Yu Pan$^{3*}$}
\address{ $^{1}$Institute of Theoretical Physics, Shanxi Datong University, Datong,
037009, China \\ $^{2}$School of Mathematics and Physics, Mianyang Teachers' College,
Mianyang 621000, China\\$^{3}$College of Science, Chongqing University of
Posts and Telecommunications, Chongqing, 400065, China  \emph{panyu@cqupt.edu.cn}}

\begin{abstract}
On the basis of the state parameter of de Sitter space-time satisfying the first law of thermodynamics, we can derive some effective thermodynamic quantities. When the temperature of the black hole horizon is equal to that of the cosmological horizon, we think that the effective temperature of the space-time should have the same value. Using this condition, we obtain a differential equation of the entropy of the de Sitter black hole in the higher-dimensional de Rham, Gabadadze and Tolley (dRGT) massive gravity. Solving the differential equation, we obtain the corrected entropy and effective thermodynamic quantities of the de Sitter black hole. The results show that for multi-parameter black holes, the entropy satisfied differential equation is invariable with different independent state parameters. Therefore, the entropy of higher-dimensional dS black holes in dRGT massive gravity is only a function of the position of the black hole horizon, and is independent of other state parameters. It is consistent with the corresponding entropy of the black hole horizon and the cosmological horizon. The thermodynamic quantities of self-consistent de Sitter space-time are given theoretically, and the equivalent thermodynamic quantities have the second-order phase transformation similar to AdS black hole, but unlike AdS black hole, the equivalent temperature of de Sitter space-time has a maximum value. By satisfying the requirement of thermodynamic equilibrium and stability of space-time, the conditions for the existence of dS black holes in the universe are obtained.
\end{abstract}

\maketitle

\section{Introduction}

The research of thermal properties of black holes {\bf (hereafter BH)}is one of the topics {\bf in which} theoretical physic{\bf ist} were interest. In recent years, the study of AdS space-time's thermodynamic property has gained {\bf intensive} attention
\citep{David12,Robie17,Aruna14,Hendi17a,Cai13,Cai16,Kastor09,Kastor10,Zhang15a,Zhang15b,Hendi16,Xu14,Wen14,Wilson15,Banerjee17,Banerjee11,Banerjee12,Ma17a,Ma17b,Hendi17b,Dayyani17,Zou17a,Cheng16,Mir17,Zhao14a,Zhao13}. When the cosmological constant in AdS space-time corresponds to the pressure of the general thermodynamic system, the extended first law of thermodynamics for the black hole can be obtained. Then we compare {\bf BH} state parameter to the Van der Waal's equation of state to study {\bf a? or several} kinds of critical phenomenon in AdS space-time. The critical point and critical exponent can also be obtained in AdS space-time, and the effect of various parameter{\bf s} in space-time on phase transition {\bf is} discussed. As to de Sitter (dS) space-time, because of the black hole horizon and cosmological horizon have different radiation temperature in general, the dS space-time does not satisfy {\bf the requirements of} the thermodynamic equilibrium stability, and it limits the study of dS space-time's thermodynamic property. {\bf However,} with the in-depth study of dark energy, the study of the thermodynamic properties of dS space-time has attracted more attention \citep{Brian13,Mbarek19,Simovic18,Simovic19,Hendi17c,Sekiwa06,Kubiznak16,McInerney16,Urano09,Bhattacharya13,Azreg-Ainou15a,Azreg-Ainou15b,Cai02}. Because in the early period of inflation, {\bf the} universe is a quasi-dS space-time, and the cosmological constant introduced in the study of dS space-time plays the role of vacuum energy. If the cosmological constant corresponds to dark energy, the universe will evolve into a new dS phase. In order to construct the whole history of universe evolution, we should have a clear understanding of classical and quantum properties in dS space-time{\bf .} dS space-time's entropy and temperature are two important parameters to study its property, but both of them are not conclusive at present. {\bf There are currently two kinds of viewpoints.} In reference{\bf s} \citep{Urano09,Shankaranarayanan13,Zhang13,Bhattacharya13,Bhattacharya16,Zhao14b,Kubiznak17},
 the authors think that dS space-time's entropy is the sum of the two horizons' $S=S_{+}+S_{c} $ ($S_{+}$ and $S_{c}$ represent the entropy of black hole horizon and cosmological horizon, respectively), {\bf but}, in reference \citep{Kanti17,Kubiznak17}, they think that it is the difference of two horizons' $S=S_{c}-S_{+}$. Different effective temperatures are obtained for the same de Sitter space-time due to different values of entropy. Different effective temperature's special property in higher-dimensional Schwarzschild-de Sitter space-time {\bf were analyzed in paper \citep{Kanti17} in which}, on the basis of dimensional consistency, we assume that dS space-time's entropy
is a function in the form of $F_{n}(x)$, with $x=r_{+} / r_{c}$ ($r_{+, c}$ is the place{\bf ? radius?} of black hole horizon and cosmological horizon namely). Using space-time's thermodynamic parameter satisfy the first law of thermodynamic
and the relation of it's effective temperature with two horizon's radiation temperature to build a differential equation that $F_{n}(x)$ is satisfied. Using the initial condition that dS space-time tend to be pure dS space-time when space-time's black hole horizon tend to be zero, we solve the differential equation that $F_{n}(x)$ must be satisfied, then obtain the entropy of ds space-time. Further we obtain other thermodynamic parameters such as equivalent temperature and pressure in dS space-time. To make our discussion more universal, {\bf for instance,} we take higher-dimensional dS black holes in dRGT massive gravity space-time to discuss.

Einstein's general relativity predicts that gravity is a spin-2 mass-free particle \citep{Gupta54,Weinberg65,Feynman95}. This prediction is also the latest expression of modern physics, which is of great significance to the study of astrophysics. But whether the graviton has mass or not and the problems such as cosmological constants and the late acceleration of the origin of the universe are challenges in relativity. Generally speaking, adding mass terms to gravitational background will bring various instabilities to gravitational theory. De Rham, Gabadadze and Tolley (dRGT) propose a non-linear theory of mass gravity \citep{Rham10,Rham11,Hinterbichler12}, which eliminates Boulware-Deser ghost \citep{Boulware72} by adding higher-order interaction terms. Vegh ha{\bf d} constructed a non-trivial black hole solution in a Ricci flat horizon under 4-dimensional dRGT gravity \citep{Vegh13,Adams15}.  Later, spherically symmetric solutions \citep{Nieuwenhuizen11,Brito13,Do16a}, charged black hole solutions \citep{Berezhiani12}, including its bi-gravity extension {\bf were} also proposed \citep{Babichev14,Do16b}. In addition, in \citep{Hendi17a,Cai15,Ge15,Xu15}, the solution of charged ADS black hole in high-dimensional DRGT gravity and its corresponding thermodynamics and phase structure in large canonical ensembles and canonical ensembles {\bf were} also introduced. Ge and Zou studied the relationship between dynamic instability and thermodynamic instability in dRGT gravitation\citep{Ge15,Zou17b}.

For charged de Sitter space-time and complex space-time, taking space-time entropy as the sum of two horizons' entropy, some progress{\bf es had} been made in the study of space-time effective temperature  \citep{Sekiwa06,Urano09,Guo15,Guo16,Zhao14c,Ma15}. However, the entropy correction term caused by the interaction between two horizons has not been taken into account in the discussion. In this paper, the thermodynamic
characteristics of de Sitter space-time {\bf is} discussed on the basis of considering the correlation between black hole horizon and cosmological horizon. Effective temperature and entropy of higher-dimensional dS black
holes in dRGT massive gravity (HBHRGT) space-time are obtained and
 the influence of coupling coefficients $c_{i} m^{2}$ on the effective
temperature and entropy is analyzed. By studying, we know that the space-time entropy of de Sitter is the sum of two horizon entropies and the entropy correction term of the system caused by the interaction of two horizons.  The total entropy of the system does not show the coupling coefficients $c_{i} m^{2}$ term, but it is just a function of the horizon position, which is consistent with that of the black hole horizon and the cosmological horizon. The determination of the space-time
entropy and the effective temperature of de Sitter lays a foundation for further study of the thermodynamic characteristics of the universality of de Sitter space-time.

This paper is arranged as follows, in {\bf the} second part, we will briefly introduce the black hole horizon and cosmological horizon corresponding thermodynamic parameter in the HBHRGT space-time. The conditions for
space-time charges to be satisfied when the radiation temperatures of the two horizons are equal as given. In the third part, the total space-time entropy and effective temperature of HBHRGT satisfying the first law of
thermodynamics are given on the basis of considering the correlation between the two horizons. {\bf In t}he forth part, we {\bf give} discuss{\bf ion} the conclusion. (We use the units $G=\hbar=k_{B}=c=1$)

\section{higher-dimensional dS black holes in dRGT massive gravity}

\label{sec:method}

The form of ($n+2$)-dimensional Einstein space-time metric is

\begin{equation}  \label{2.1}
d s^{2}=-f(r) d t^{2}+f^{-1}(r) d r^{2}+r^{2} h_{ij} d x^{i} d x^{j},
\end{equation}
where $h_{ij} d x^{i} d x^{j}$ is the line element including constant curvature $\frac{n(n-1)}{k}$. $k=0$ represents a flat universe, however, if $k= \pm1$, the universe is non-flat, $k=1$  for positive curvature and $k=-1$ for negative curvature respectively.

By using the reference metric
\begin{equation}
f_{\mu \nu }=\operatorname{diag} \left( 0,0,c_{0}^{2}h_{ij}\right),   \label{2.2}
\end{equation}%
the metric function $f(r)$ is obtained as \citep{Cai15,Zou17b},
\begin{equation}
\begin{aligned} &f(r)=k+c_{0}^{2} c_{2} m^{2}-\frac{m_{0}}{r^{n-1}}-\frac{2
\Lambda r^{2}}{n(n+1)}+\frac{(n-1) c_{0}^{3} c_{3} m^{2}}{r} \\
&+\frac{c_{0} c_{1} m^{2}}{n} r+\frac{(n-1)(n-2) c_{0}^{4} c_{4}
m^{2}}{r^{2}}+\frac{q^{2}}{2 n(n-1) r^{2(n-1)}}. \end{aligned}  \label{2.3}
\end{equation}
where $\Sigma_{n}$ is the volume of space spanned by coordinates $x^{i}$,  $c_{0}$ is a positive constant, and $m$ is {\bf the mass of} the black hole. {\bf We note} that the terms $c_{3} m^{2}$ and  $c_{4} m^{2}$ only appear  for $n \geq 3$ and $n \geq 4$ in the black hole solutions, respectively \citep{Cai15}. When $m \rightarrow 0$, namely£¬ there is no mass, Eq. (\ref{2.3}) goes back to the $(n+2)$-dimensional Schwarzschild dS (AdS) black hole solution.

When cosmological constants $\Lambda<0$, the spacetime is named AdS and there is only one black hole
horizon. As shown in reference \citep{Cai15,Zou17b} there are some discussions about black
hole phase translation and critical phenomena. When $\Lambda>0$, it is dS
space-time, there are not only black hole horizon $r_{+}$, but also
cosmological horizon $r_{c}$. The thermodynamic parameters of the two
horizon satisfy the first law of thermodynamics \citep{Brian13,Sekiwa06,Kubiznak16,Gunasekaran12}.

Some thermodynamic quantities associated with the cosmological horizon are
\begin{equation}  \label{2.2a}
\begin{aligned} Q=&\frac{\Sigma_{n}q}{16 \pi},S_{c}=\frac{\Sigma_{n}}{4}
r_{c}^{n}, V_{c}=\frac{\Sigma_{n}}{n+1} r_{c}^{n+1}, P=-\frac{\Lambda}{8
\pi}<0, \\ T_{c}=&-\frac{1}{4 \pi r_{c}}[(n-1) k-\frac{2 r_{c}^{2}
\Lambda}{n}-\frac{q^{2}}{2 n r_{c}^{2(n-1)}}\\ &+(n-1) c_{2} c_{0}^{2}
m^{2}+\frac{(n-1)(n-2) c_{3} c_{0}^{3} m^{2}}{r_{c}}\\ &+c_{1} c_{0} m^{2}
r_{c}+\frac{(n-1)(n-2)(n-3) c_{4} c_{0}^{4} m^{2}}{r_{c}^{2}}]. \end{aligned}
\end{equation}

Furthermore, the first law of thermodynamics of the cosmological horizon is %
\citep{Brian13,Sekiwa06,Kubiznak16}
\begin{equation}  \label{2.3a}
\begin{split}
d M=&-T_{c} d S_{c}+U_{c} d Q+V_{c} d P \\
&+C_{1 c} d c_{1}+C_{2 c} d c_{2}+ C_{3 c} d c_{3}+C_{4 c} d c_{4},
\end{split}%
\end{equation}
where $M=\frac{n \Sigma_{n}}{16 \pi} m_{0}$.

The corresponding potentials are
\begin{equation}  \label{2.4}
\begin{aligned} U_{c}&=\frac{q}{(n-1) r_{c}^{n-1}}, \quad C_{1c}=\frac{c_{0}
m^{2} \Sigma_{n} r_{c}^{n}}{16 \pi},\\ C_{2c}&=\frac{n c_{0}^{2} m^{2}
\Sigma_{n} r_{c}^{n-1}}{16 \pi}, C_{3 c}=\frac{n(n-1) c_{0}^{3} m^{2}
\Sigma_{n} r_{c}^{n-2}}{16 \pi},\\ C_{4 c}&=\frac{n(n-1)(n-2) c_{0}^{4} m^{2}
\sum_{n} r_{c}^{n-3}}{16 \pi}. \end{aligned}
\end{equation}

For the black hole horizon, {\bf the }associated thermodynamic quantities are
\begin{equation}  \label{2.5}
\begin{aligned} S_{+}=&\frac{\Sigma_{n}}{4} r_{+}^{n}, \quad
V_{+}=\frac{\sum_{n}}{n+1} r_{+}^{n+1}, \\ T_{+}=&\frac{1}{4 \pi
r_{+}}[(n-1) k-\frac{2 r_{+}^{2} \Lambda}{n}-\frac{q^{2}}{2 n
r_{+}^{2(n-1)}}+c_{1} c_{0} m^{2} r_{+}\\ &+(n-1) c_{2} c_{0}^{2}
m^{2}+\frac{(n-1)(n-2) c_{3} c_{0}^{3} m^{2}}{r_{+}}\\
&+\frac{(n-1)(n-2)(n-3) c_{4} c_{0}^{4} m^{2}}{r_{+}^{2}}]. \end{aligned}
\end{equation}

The thermodynamic quantities of black hole horizon satisfy the first law,
\begin{equation}  \label{2.6}
\begin{aligned} d M=&T_{+} d S_{+}+U_{+} d Q+V_{+} d P\\ &+C_{1+} d
c_{1}+C_{2+} d c_{2}+C_{3+} d c_{3}+C_{4+} d c_{4}, \end{aligned}
\end{equation}
where the corresponding potential and parameter $C_{i}$ respectively
are
\begin{equation}  \label{2.7}
\begin{aligned} U_{+}&=\frac{q}{(n-1) r_{+}^{n-1}}, C_{1+}=\frac{c_{0} m^{2}
\Sigma_{n} r_{+}^{n}}{16 \pi},\\ C_{2+}&=\frac{n c_{0}^{2} m^{2} \Sigma_{n}
r_{+}^{n-1}}{16 \pi}, C_{3+}=\frac{n(n-1) c_{0}^{3} m^{2} \Sigma_{n}
r_{+}^{n-2}}{16 \pi},\\ C_{4+}&=\frac{n(n-1)(n-2) c_{0}^{4} m^{2} \Sigma_{n}
r_{+}^{n-3}}{16 \pi}, \end{aligned}
\end{equation}
from horizon's equation $f\left(r_{+, c}\right)=0$, we get that
\begin{equation}  \label{2.8}
\begin{aligned} \frac{2
\Lambda}{n(n+1)}=&-\frac{q^{2}\left(1-x^{n-1}\right)}{2 n(n-1) r_{c}^{2 n}
x^{n-1}\left(1-x^{n+1}\right)}\\ &+\frac{c_{0} c_{1} m^{2}}{n}
\frac{\left(1-x^{n}\right)}{r_{c}\left(1-x^{n+1}\right)}\\
&+\left(k+c_{0}^{2} c_{2} m^{2}\right)
\frac{\left(1-x^{n-1}\right)}{r_{c}^{2}\left(1-x^{n+1}\right)}%
\\&+(n-1)c_{0}^{3} c_{3} m^{2}
\frac{\left(1-x^{n-2}\right)}{r_{c}^{3}\left(1-x^{n+1}\right)}\\&+(n-1)(n-2)
c_{0}^{4} c_{4} m^{2}
\frac{\left(1-x^{n-3}\right)}{r_{c}^{4}\left(1-x^{n+1}\right)}, \end{aligned}
\end{equation}
\begin{equation}  \label{2.9}
\begin{aligned} \frac{16\pi M}{n \Sigma_{n}}=&\left(k+c_{0}^{2} c_{2}
m^{2}\right) r_{c}^{n-1} x^{n-1}
\frac{\left(1-x^{2}\right)}{\left(1-x^{n+1}\right)}\\
&+\frac{q^{2}\left(1-x^{2 n}\right)}{2 n(n-1) r_{c}^{n-1}
x^{n-1}\left(1-x^{n+1}\right)}\\ &+\frac{c_{0} c_{1} m^{2}}{n} r_{c}^{n}
x^{n} \frac{(1-x)}{\left(1-x^{n+1}\right)}\\ &+(n-1) c_{0}^{3} c_{3} m^{2}
r_{c}^{n-2} x^{n-2} \frac{\left(1-x^{3}\right)}{\left(1-x^{n+1}\right)} \\
&+(n-1)(n-2) c_{0}^{4} c_{4} m^{2} r_{c}^{n-3} x^{n-3}
\frac{\left(1-x^{4}\right)}{\left(1-x^{n+1}\right)}. \end{aligned}
\end{equation}
where $x=r_{+}/r_{c}$.

When $\tilde{T}_{+}=\tilde{T}_{c}=T$, from Eq. (\ref{2.2a}) and Eq. (\ref{2.5}) we get
\begin{equation}  \label{2.10}
\begin{aligned} \frac{2 \Lambda}{n}=&\frac{(n-1)}{r_{c}^{2} x}\left(k+c_{2}
c_{0}^{2} m^{2}\right)\\ &-\frac{q^{2}}{2 n r_{c}^{2 n} x^{2 n-1}}
\frac{1+x^{2 n-1}}{(1+x)}+2 \frac{c_{1} c_{0} m^{2}}{r_{c}(1+x)}\\
&+(n-1)(n-2) c_{3} c_{0}^{3} m^{2}
\frac{\left(1+x^{2}\right)}{r_{c}^{3}x^{2}(1+x)}\\&+\frac{(n-1)(n-2)(n-3)
c_{4} c_{0}^{4} m^{2}}{r_{c}^{4} x^{3}} \frac{\left(1+x^{3}\right)}{(1+x)},
\end{aligned}
\end{equation}
From Eq. (\ref{2.8}) and Eq. (\ref{2.10}), we get a relevant expression among the parameters of $\frac{q^{2}}{r_{c}^{2 n-2}}, \left(k+c_{0}^{2} c_{2}
m^{2}\right), c_{0} c_{1} m^{2} r_{c}, \frac{c_{0}^{3} c_{3} m^{2}}{r_{c}}$
and $\frac{c_{0}^{4} c_{4} m^{2}}{r_{c}^{2}}$ , when black hole
horizon and cosmological horizon have the same radiation temperature.

\begin{equation}  \label{2.11}
\setlength{\abovedisplayskip}{1pt} \setlength{\belowdisplayskip}{3pt} %
\begin{aligned} \frac{q^{2}}{r_{c}^{2 n-2}} K(x, n)=&\left(k+c_{0}^{2} c_{2}
m^{2}\right) A(x, n)+c_{0} c_{1} m^{2} r_{c} B(x, n)\\ &+\frac{c_{0}^{3}
c_{3} m^{2}}{r_{c}} C(x, n)+\frac{c_{0}^{4} c_{4} m^{2}}{r_{c}^{2}} D(x, n),
\end{aligned}
\end{equation}
where
\begin{equation}  \label{2.12}
\begin{aligned}
K(x,n)=&\left(1+x^{n}\right)\left(1-x^{2n}\right)-n(1-x^{n})\\
 &[1+x^{2n}+2n x^{n+1}\left(1-x^{n-2}\right)], \\
A(x,n)=&2n(n-1)x^{2n-2}\\&[(1+x)^{2}\left(1-x^{n}\right)-n\left(1-x^{2}\right)\left(1+x^{n}\right)],\\
B(x, n)=&2(n-1) x^{2n-1}\\&[\left(1-x^{n}\right)(1+x)-n(1-x)\left(1+x^{n}\right)], \\
C(x, n)=&2n(n-1)^{2} x^{2 n-3}[2+3 x^{2}-x^{n}+x^{3}\\
 &-3 x^{n+1}-2x^{n+3}-n\left(1+x^{n}\right)\left(1-x^{3}\right)], \\
D(x, n)=&2n(n-1)^{2}(n-2) x^{2 n-4}[3+4 x^{3}-x^{n}+x^{4}\\
 &-4 x^{n+1}-3x^{n+4}-n\left(1+x^{n}\right)\left(1-x^{4}\right)]. \end{aligned}
\end{equation}

From Eq. (\ref{2.11}) we know {\bf that} $\left(k+c_{0}^{2} c_{2} m^{2}\right)$, $\frac{%
q^{2}}{r_{c}^{2 n-2}}$, $\frac{c_{0}^{3} c_{3} m^{2}}{r_{c}}$, $\frac{%
c_{0}^{4} c_{4} m^{2}}{r_{c}^{2}}$ and $c_{0} c_{1} m^{2} r_{c}$ are not
completely independent, when black hole horizon and cosmological horizon have the same
radiation temperature, any of these variables can be expressed as functions
of other variables. When we use $\frac{q^{2}}{r_{c}^{2 n-2}}$ as other
variables' function , substituting Eq. (\ref{2.10}) and Eq. (\ref{2.11}) into Eq. (%
\ref{2.2a}) or Eq. (\ref{2.5}), we obtain the temperature when two horizons
have same radiation temperature.
\begin{equation}  \label{2.13}
\begin{aligned} T_{K}=&T_{+, k}=T_{c, k} \\ =&\frac{2\left(k+c_{2} c_{0}^{2}
m^{2}\right)}{4 \pi r_{c} K(x, n)}\\ &\quad [n^{2}
x^{n-1}(1-x)\left(1-x^{n-1}\right)-n\left(1-x^{n+1}\right)\\ &\quad
\left(1-x^{2 n-2}\right)+\left(1-x^{2 n}\right)\left(1-x^{n-1}\right)] \\
&+\frac{c_{1} c_{0} m^{2}}{4 \pi n K(x, n)}\\ &\quad [2 n^{2}
x^{n}(1-x)\left(1-x^{n-1}\right)-n\left(1-x^{2 n}\right)\\ &\quad
\left(1-x^{n}\right)+\left(1-x^{2 n}\right)\left(1-x^{n}\right)]\\
&+\frac{(n-1) c_{3} c_{0}^{3} m^{2}}{4 \pi r_{c}^{2} K(x, n)}\\
&\quad[2n^{2}x^{n-2}\left(1-x^{3}\right)(1-x^{n-1})\\ &\quad
+3\left(1-x^{n-2}\right)\left(1-x^{2n}\right)-n\\ &\quad (3+3 x^{3 n-2}-4
x^{n+1}-4 x^{2 n-3}+x^{n-2}+x^{2 n})]\\ &+\frac{2(n-1)(n-2) c_{4} c_{0}^{4}
m^{2}}{4 \pi r_{c}^{3} K(x, n)}\\ &\quad [2\left(1-x^{n-3}\right)
\left(1-x^{2 n}\right)\\ &\quad
+n^{2}x^{n-3}\left(1-x^{n-1}\right)\left(1-x^{4}\right)-n\\ &\quad
(2+x^{n-3}+x^{2 n}+2 x^{3 n-3}-3 x^{2 n-4}-3 x^{n+1})]. \end{aligned}
\end{equation}
When we select $c_{0} c_{1} m^{2} r_{c}$ as the function of other variables, we
substitute Eq. (\ref{2.10}) and Eq. (\ref{2.11}) into Eq. (\ref{2.2a}) or Eq. (\ref{2.5}) to obtain the temperature when the radiation temperature of two
horizons is equal.
\begin{equation}  \label{2.14}
\begin{aligned} 4 \pi r_{c}& B(x, n) T_{B}=4 \pi r_{c} B(x, n) T_{+, B}=4
\pi r_{c} B(x, n) T_{c, B}\\ &=-2(n-1)\left(k+c_{2} c_{0}^{2} m^{2}\right)
x^{2 n-2}(1-x)\\
&\qquad[\left(1-x^{n}\right)(1+x)-n(1-x)\left(1+x^{n}\right)]\\ &\quad- 2
\frac{c_{4} c_{0}^{4} m^{2}}{r_{c}^{2}}(n-1)^{2}(n-2) x^{2 n-4}\\
&\qquad[3\left(1-x^{4}\right)\left(1-x^{n}\right)-n\\ &\qquad\left(1+3
x^{n}+3 x^{4}-4 x^{3}-4 x^{n+1}+x^{n+4}\right)]\\ &\quad+\frac{q^{2}}{n
r_{c}^{2 n-2}}\\ &\qquad [\left(1-x^{n}\right)\left(1-x^{2
n}\right)-n\left(1-x^{n}\right)\\ &\qquad\left(1-x^{2 n}\right)+2 n^{2}
x^{n}(1-x)\left(1+x^{n-1}\right)]\\ &\quad-\frac{2 c_{3} c_{0}^{3}
m^{2}}{r_{c}}(n-1)^{2} x^{2n-3}(1-x)\\
&\qquad[2\left(1+x+x^{2}\right)\left(1-x^{n}\right)\\
&\qquad-n(1-x)\left(1+2 x+2 x^{n}+x^{n+1}\right)], \end{aligned}
\end{equation}
when we take $\left(k+c_{0}^{2} c_{2} m^{2}\right)$ as other functions,
from Eq. (\ref{2.10}), Eq. (\ref{2.11}), Eq. (\ref{2.2a}) and Eq. (\ref{2.5}), we can obtain the temperature when two horizons have same radiation temperature.
\begin{equation}  \label{2.15}
\begin{aligned} &4\pi r_{c} A(x, n) T_{A}= 4 \pi r_{c} A(x, n) T_{+, A}=4
\pi r_{c} A(x, n) T_{c, A} \\ &=-\frac{q^{2}}{r_{c}^{2 n-2}} 2(n-1)\\
&\qquad\left[\left(1-x^{n-1}\right)\left(1-x^{2 n}\right)-n
x^{n-1}(1-x)\left(1-x^{n}\right)\right]\\ &\quad+2 c_{1} c_{0} m^{2}
r_{c}(n-1)(1-x) x^{2 n-2}\\
&\qquad[\left(1-x^{n}\right)(1+x)-n(1-x)\left(1+x^{n}\right)]\\ &\quad -2
\frac{c_{0}^{3} c_{3} m^{2}}{r_{c}} n(n-1)^{2} x^{2 n-3}\\ &\qquad
[\left(4-3 x-x^{3}-x^{n-1}-3 x^{n+1}+4 x^{n+2}\right)-n\\ &\qquad \left(2-3
x+x^{3}+x^{n-1}-3 x^{n+1}+2 x^{n+2}\right)]\\ &\quad -2 \frac{c_{0}^{4}
c_{4} m^{2}}{r_{c}^{2}} n(n-1)^{2}(n-2) x^{2 n-5}\\ &\qquad [x\left(6-4
x^{2}-2 x^{4}-2 x^{n-1}-4 x^{n+1}+6 x^{n+3}\right)+n \\ &\qquad (2+4 x^{3}-3
x^{4}-x^{5}-4 x^{n}-2 x^{n+1}+4 x^{n+2}\\ &\qquad-3 x^{n+4}+3 x^{n+5})],
\end{aligned}
\end{equation}
when we take $\frac{c_{0}^{3} c_{3} m^{2}}{r_{c}}$ as other functions, substituting Eq. (%
\ref{2.10}) and Eq. (\ref{2.11}) into Eq. (\ref{2.2a}) or Eq. (\ref{2.5}), when two horizons have same radiation temperature, the temperature is as following .
\begin{equation}  \label{2.16}
\begin{aligned}
 &4\pi r_{c} C(x, n) T_{c}=4 \pi r_{c} C(x, n) T_{+, c}=4 \pi r_{c} C(x, n) T_{c, c} \\
 &=2 n(n-1)^{2}\left(k+c_{2} c_{0}^{2} m^{2}\right) x^{2 n-3}\\
 &\qquad[\left(4-3 x-x^{3}-x^{n-1}-3 x^{n+1}+4 x^{n+2}\right)-n\\
 &\qquad(2-3 x+x^{3}+x^{n+1}-3 x^{n+1}+2 x^{n+2})]\\
 &\quad-\frac{q^{2}(n-1)}{r_{c}^{2 n-2} x^{2}}\\
 &\qquad[-3 x^{2}\left(1-x^{2n}\right)\left(1-3 x^{n-2}\right)+n \\
 &\qquad\left(3 x^{2}+x^{n}+3 x^{3n}+x^{2 n+2}-4 x^{n+3}-4 x^{2 n-1}\right)\\
 &\qquad-2 n^{2}x^{n}\left(1-x^{3}\right)\left(1-x^{n-1}\right)]\\
 &\quad+2 c_{1} c_{0}m^{2} r_{c}(n-1)^{2} x^{2 n-3}\\
 &\qquad[2\left(1-x^{3}\right)\left(1-x^{n}\right)+n\\
 &\qquad\left(-1+3x^{2}-2 x^{3}-2 x^{n}+3 x^{n+1}-x^{n+3}\right)] \\
 &\quad+\frac{c_{4}c_{0}^{4} m^{2}}{r_{c}^{2}} 2 n(n-1)^{3}(n-2) x^{2 n-6}\\
 &\qquad[\left(x^{n}-9 x^{2}+8 x^{3}+x^{6}+8 x^{n+3}-9 x^{n+4}\right)+n \\
 &\qquad\left(x^{n}+x^{6}-4 x^{3}+3 x^{2}-4 x^{n+3}+3 x^{n+4}\right)],
\end{aligned}
\end{equation}

When we take $\frac{c_{0}^{4} c_{4} m^{2}}{r_{c}^{2}}$ as other functions, we substitute Eq. (\ref{2.10}) and Eq. (\ref{2.11}) into Eq. (\ref{2.2a}%
) or Eq. (\ref{2.5}) to obtain the temperature when the radiation
temperature of two horizons is equal.

\begin{equation}  \label{2.17}
\begin{aligned} &4\pi r_{c} D(x, n) T_{d}=4 \pi r_{c} D(x, n) T_{+, d}=4 \pi r_{c} D(x, n) T_{c, d} \\
 &=\left( k+c_{2} c_{0}^{2} m^{2}\right) 2n(n-1)^{3} (n-2) x^{2 n-5}\\
 &\qquad[\left(6-6 x^{4}+4 x^{3}-4 x^{n}-6 x^{n+1}-6 x^{5}+4 x^{n+2}\right.\\
 &\qquad\left.+6 x^{n+6}\right)-n\left(1-x^{4}\right)\left(5-3 x+3 x^{n}-5 x^{n+1}\right)\\
 &\qquad+n^{2}\left(1-x^{4}\right)(1-x)\left(1+x^{n}\right)] \\
 &\quad-\frac{q^{2}}{r_{c}^{2 n-2}} 2(n-1)(n-2)[-2\left(1-x^{2 n}\right)\left(1-x^{n}\right) \\
 &\qquad +n x\left(1+x^{n-4}-3 x^{n}-3 x^{2n-5}+x^{2 n-1}+2 x^{3 n-4}\right)\\
 &\qquad -n^{2}x^{n-3}\left(1-x^{4}\right)\left(1-x^{n-1}\right)]\\
 &\quad+2(n-1)^2(n-2)c_1c_0m^2r_c^2x^{2n-4}[3(1-x^4)(1-x^n)\\
 &\qquad +n\left(-1+4 x^{3}-3 x^{4}+4 x^{n+1}-3 x^{n}-x^{n+4}\right)] \\
 &\quad+2 n(n-1)^{3}(n-2) \frac{c_{3} c_{0}^{3} m^{2}}{r_{c}} x^{2 n-6}\\
 &\qquad[\left(9 x^{2}-8 x^{3}-6 x^{6}-8 x^{n+3}+9 x^{n+4}\right)+n \\
 &\qquad \left(-3 x^{2}-4 x^{3}-x^{n}-x^{6}+4 x^{n+3}-3 x^{n+4}\right)].
\end{aligned}
\end{equation}

The temperature of two horizons with the same radiation temperature
expressed by different independent variables is given by Eqs. (\ref{2.13})-(%
\ref{2.17}).

\section{Effective thermodynamic quantity}

\label{sec:result} We regard the HBHRGT space-time as a thermodynamic
system, for which the state parameters satisfy the {\bf f}irst law of thermodynamics.
Considering the connection between the black hole horizon and the
cosmological horizon, we can derive the effective thermodynamic quantities
and the corresponding first law of black hole thermodynamics
\begin{equation}  \label{3.1}
\begin{aligned} d M= & T_{eff} d S-P_{eff} d V+\phi_{eff} d Q \\ &+C_{1} d
c_{1}+ C_{2} d c_{2}+C_{3} d c_{3}+C_{4} d c_{4}, \end{aligned}
\end{equation}

Here the thermodynamic volume is that between the black hole horizon and the
cosmological horizon, namely \citep{Brian13}
\begin{equation}
V=V_{c}-V_{+}=\frac{\sum_{n}}{n+1}r_{c}^{n+1}\left( 1-x^{n+1}\right),
\label{3.2}
\end{equation}

Considering that black hole horizon and cosmological horizon are not independent, the
entropy is \citep{Zhao14b,Li17}
\begin{equation}\label{3.3}
S=S_{c}+S_{+}+S_{t}=\frac{\Sigma_{n}}{4} r_{c}^{n}\left(1+x^{n}+f_{n}(x)\right)\\
=\frac{\Sigma_{n}}{4} r_{c}^{n} F_{n}(x),
\end{equation}

Here the undefined function $f_{n}(x)$ represents the extra contribution
from the correlations of the two horizons. From (\ref{3.1}), the system's
effective temperature $T_{eff}$, pressure $p_{eff}$ and potential $\phi_{eff}$,
respectively can be represented as
\begin{equation}  \label{3.4}
\begin{aligned} T_{eff}&=\left(\frac{\partial M}{\partial S}\right)_{Q,V,
c_{i}}\\ &=\frac{\left(\frac{\partial M}{\partial
x}\right)_{r_{c}}\left(\frac{\partial V}{\partial r_{c}}\right)_{x}
-(\frac{\partial V}{\partial x})_{r_c}(\frac{\partial M}{\partial
r_{c}})_{x}}{(\frac{\partial S}{\partial x})_{r_{c}}(\frac{\partial
V}{\partial r_{c}})_{x}-(\frac{\partial V}{\partial
x})_{r_{c}}(\frac{\partial {S}}{\partial r_{c}})_{x}}, \end{aligned}
\end{equation}

\begin{equation}  \label{3.5}
\begin{aligned} P_{eff}&=-\left(\frac{\partial M}{\partial V}\right)_{Q,S,
c_{i}}\\ &=\frac{\left(\frac{\partial S}{\partial
x}\right)_{r_{c}}\left(\frac{\partial M}{\partial
r_{c}}\right)_{x}-\left(\frac{\partial M}{\partial
x}\right)_{r_{c}}\left(\frac{\partial S}{\partial
r_{c}}\right)_{x}}{\left(\frac{\partial V}{\partial
x}\right)_{r_{c}}\left(\frac{\partial S}{\partial r_{c}}\right)_{x}
-\left(\frac{\partial S}{\partial x}\right)_{r_{c}}\left(\frac{\partial
V}{\partial r_{c}}\right)_{x}}, \end{aligned}
\end{equation}

From Eq. (\ref{2.9}), Eq. (\ref{3.2}) and Eq. (\ref{3.3}), $T_{eff}$ can be expressed as

\begin{equation}  \label{3.6}
T_{eff}=\frac{n B(x, q)}{4 \pi r_{c} A(x)},
\end{equation}
Where
\begin{equation}  \label{3.7}
\begin{aligned} B&(x, q)=\left(k+c_{0}^{2} c_{2} m^{2}\right) x^{n-2}\\
&\qquad\frac{n\left(1-x^{2}\right)\left(1+x^{n+1}\right)-\left(1+x^{2}%
\right)\left(1-x^{n+1}\right)}{\left(1-x^{n+1}\right)}\\
&\quad+\frac{q^{2}\left[\left(1-x^{2n}\right)\left(1+x^{n+1}\right)-n%
\left(1+x^{2n}\right)\left(1-x^{n+1}\right)\right]}{2 n(n-1) r_{c}^{2 n-2}
x^{n}\left(1-x^{n+1}\right)}\\ &\quad+\frac{c_{0} c_{1} m^{2}}{n} r_{c}
x^{n-1}
\frac{n(1-x)\left(1+x^{n+1}\right)-x\left(1-x^{n}\right)}{\left(1-x^{n+1}%
\right)}\\ &\quad+(n-1)c_0^3c_3m^2x^(n-4)\\
&\qquad\frac{n(1-x^3)(1+x^{n+1})-(2+x^3-x^{n+1}-2x^{n+4})}{r_c(1-x^{n+1})}\\
&\quad+(n-1)(n-2)c_0^4c_4m^2x^{n-4}\\
&\qquad\frac{n(1-x^4)(1+x^{n+1})-(3+x^4-x^{n+1})-3x^{n+5}}{r_c^2(1-x^{n+1})},
\end{aligned}
\end{equation}

\begin{equation}
\begin{aligned}\label{3.8} A(x)&=\left[n
x^{n-1}+f_{n}^{\prime}(x)\right]\left[1-x^{n+1}\right]\\ &\qquad+n
x^{n}\left[1+x^{n}+f_{n}(x)\right]\\ &=\left[1-x^{n+1}\right]
F_{n}^{\prime}(x)+n x^{n} F_{n}(x). \end{aligned}
\end{equation}

The space-time's effective temperature should equal to radiation temperature, when the two
horizon{\bf s} have the same radiation temperature, that is
\begin{equation}  \label{3.9}
\tilde{T}_{eff}=T=\tilde{T}_{+}=\tilde{T}_{c},
\end{equation}

Substituting (\ref{2.11}) into (\ref{3.7}), from (\ref{3.6}) we get
\begin{equation}  \label{3.10}
A(x)=\frac{n \tilde{B}(x)}{4 \pi r_{c} \tilde{T}_{eff}},
\end{equation}
when $\frac{q^{2}}{r_{c}^{2 n-2}}$ is other variables' function , the $%
\tilde{T}_{eff}=T_{k}$ can be obtained from Eq. (\ref{2.13}), and $\tilde{B}(x)$ can
be shown
\begin{equation}  \label{3.11}
\begin{aligned} \tilde{B}&(x) \left(1-x^{n+1}\right)=\frac{2(k+c_{0}^{2}
c_{2} m^{2})}{K(x, n)}x^{n-1}(1+x^{n+2})\\
&\qquad[n^{2}x^{n-1}(1-x^{2})(1-x^{n-1})-n(1-x^{n+1})\\ &\qquad(1-x^{2
n-2})+(1-x^{2 n})(1-x^{n-1})]\\ &\quad+\frac{c_{0} c_{1} m^{2} r_{c}}{n K(x,
n)} x^{n-1}\left(1+x^{n+2}\right)\\ &\qquad[2 n^{2}
x^{n}(1-x)\left(1-x^{n-1}\right)-n\left(1-x^{n}\right)\\ &\qquad\left(1-x^{2
n}\right)+\left(1-x^{n}\right)\left(1-x^{2 n}\right)]\\ &\quad+(n-1)
c_{0}^{3} c_{3} m^{2} x^{n-1} \frac{\left(1+x^{n+2}\right)}{r_{c} K(x, n)}\\
&\qquad\left[2 n^{2}
x^{n-2}\left(1-x^{3}\right)\left(1-x^{n-1}\right)\right.\\
&\qquad+3\left(1-x^{2 n}\right)\left(1-x^{n-2}\right)-n\\ &\qquad \left(3+3
x^{3 n-2}-4 x^{n+1}-4 x^{2 n-3}-x^{n-2}+x^{2 n}\right) ]\\
&\quad+\frac{2(n-1)(n-2) c_{0}^{4} c_{4} m^{2}
x^{n-1}\left(1+x^{n+2}\right)}{r_{c}^{2} K(x, n)}\\ &\qquad[2(1-x^{2
n})(1-x^{n-3})-n\\ &\qquad(2+x^{n-3}+x^{2 n}+2 x^{3 n-3}-3 x^{2 n-4}-3
x^{n+1})\\ &\qquad +n^{2} x^{n-3}(1-x^{n-1})(1-x^{4})].
\end{aligned}
\end{equation}

Putting Eq. (\ref{2.13}) and Eq. (\ref{3.11}) into Eq. (\ref{3.10}) we get
\begin{equation}  \label{3.12}
A(x)=\frac{n x^{n-1}\left(1+x^{n+2}\right)}{\left(1-x^{n+1}\right)}.
\end{equation}

The different results of $\tilde{B}(x)$  are obtained when we select different state parameters. By substituting $\tilde{B}(x)$ and the corresponding radiation
temperature $T=\tilde{T}_{+}=\tilde{T}_{c}$ into Eq. (\ref{3.10}), the
same results are obtained as Eq. (\ref{3.12}). It is shown that Eq. (%
\ref{3.12}) has nothing to do with the selected variables and is a universal
relationship between space and time. Solving Eq. (\ref{3.12}), we get
\begin{equation}  \label{3.13}
\begin{aligned} F_{n}(x)&=\frac{3 n+2}{2
n+1}\left(1-x^{n+1}\right)^{\frac{n}{n+1}}\\ &\quad-\frac{(n+1)\left(1+x^{2
n+1}\right)-(2 n+1) x^{n}(1+x)}{(2 n+1)\left(1-x^{n+1}\right)}\\ &=\frac{3
n+2}{2 n+1}\left(1-x^{n+1}\right)^{\frac{n}{n+1}}-\frac{(n+1)\left(1+x^{2
n+1}\right)}{(2 n+1)\left(1-x^{n+1}\right)}\\ &\quad-\frac{\left(1-2
x^{n+1}-x^{2 n+1}\right)}{\left(1-x^{n+1}\right)}+1+x^{n}\\
&=f_{n}(x)+1+x^{n}, \end{aligned}
\end{equation}
when we solve Eq. (\ref{3.12}), we take $F_{n}(0)=1$ and $f_{n}(0)=0$. It is
considering that when $x \rightarrow 0$, $r_{+}<<r_{c}$ is just that
space-time tend{\bf s} to be pure dS space-time. As for different space dimension $%
n=2$, $n=3$, {\bf and} $n=4$, Eq. (\ref{3.13}) can be written as
\begin{equation}  \label{3.14}
\begin{aligned}
F_{n=2}(x)&=\frac{8}{5}(1-x^{3})^{\frac{2}{3}}-\frac{3\left(1+x^{5}\right)-5
x^{2}(1+x)}{5\left(1-x^{3}\right)}\\
&=\frac{8}{5}(1-x^{3})^{\frac{2}{3}}-\frac{2\left(4-5
x^{3}-x^{5}\right)}{5\left(1-x^{3}\right)}+1+x^{2}\\ &=f_{n=2}(x)+1+x^{2},\\
F_{n=3}(x)&=\frac{11}{7}\left(1-x^{4}\right)^{\frac{3}{4}}-\frac{4%
\left(1+x^{7}\right)-7 x^{3}(1+x)}{7\left(1-x^{4}\right)} \\
&=\frac{11}{7}\left(1-x^{4}\right)^{3 / 4}-\frac{11-14 x^{4}-3
x^{7}}{7\left(1-x^{4}\right)}+1+x^{3}\\ &=f_{n=3}(x)+1+x^{3},\\ F_{n=4}(x)
&=\frac{14}{9}\left(1-x^{5}\right)^{\frac{4}{5}}-\frac{5\left(1+x^{9}%
\right)-9 x^{4}(1+x)}{9\left(1-x^{5}\right)} \\
&=\frac{14}{9}\left(1-x^{5}\right)^{\frac{4}{5}}-\frac{2\left(7-9 x^{5}-2
x^{9}\right)}{9\left(1-x^{5}\right)}+1+x^{4}\\ &=f_{n-4}(x)+1+x^{4}.
\end{aligned}
\end{equation}

\begin{figure}[htp]
\begin{minipage}[t]{0.45\textwidth}
  \centering
  \includegraphics[width=3in]{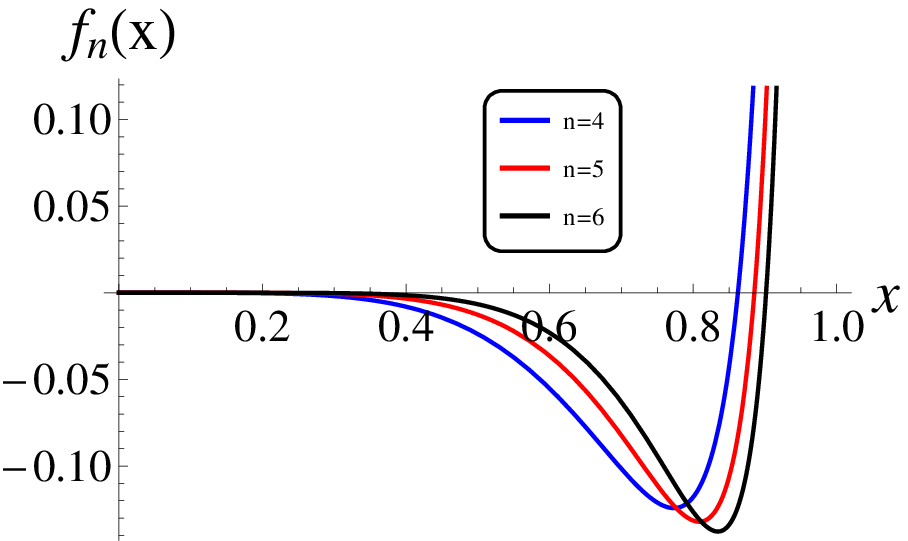}
  \caption{$ f(x)-x $ diagram for diffrent n.}\label{fig1}
\end{minipage}
\begin{minipage}[t]{0.36\textwidth}
  \centering
  \includegraphics[width=2.8in]{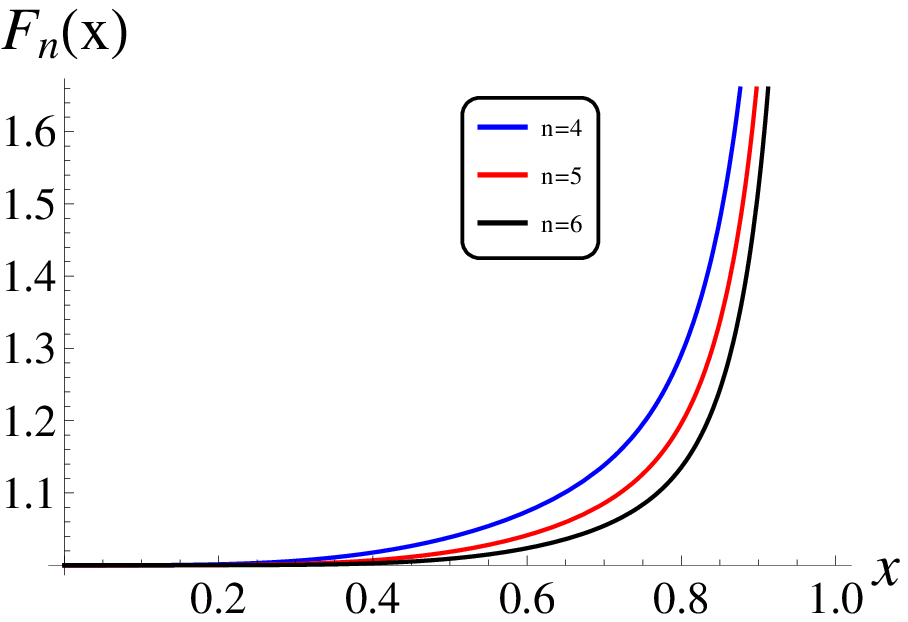}
  \caption{$ F(x)-x $ diagram for diffrent n.}\label{fig2}
\end{minipage}
\end{figure}

Meanwhile, we can find the zero position of $f_{n}(x)-x$, and the $x_0$ when
$f_{n}\left(x_{0}\right)=0$. According to Eq. (\ref{3.13}), the interaction terms $f_{n}(x)$ are
different in different dimensions, but the general trend is the same. At the point $x=x_0$, the correction term of entropy is vanished. When $x_{0}<x<1$, the correction term of entropy has the positive value and increases with the increasing of x. At that time, it tends to be infinite when $x \rightarrow 1$. When the correction term of interval entropy is negative in the region $0<x<x_{0}$ and has a minimum value $x=x_{m}$, From Fig. \ref{fig1} we know that both $x_{0}$ and $x_{m}$ increase as the dimension $n$ increases. The results show that with the increase of dimension $n$, the region in which the corrected entropy increases, will decrease, while {\bf the other region,}  will increase. Meanwhile the modified value of entropy is a function of dimension.

In the same way, we choose other independent variables to
discuss, we all get the differential Eq. (\ref{3.12}) that space-time
entropy must be satisfied in different independent variables. From Eq. (\ref{3.12}) we
know that the space-time entropy's correction term $f(x)$ only relate to black hole
horizon and cosmological horizon's place. This point match{\bf s} to black hole
horizon and cosmological horizon corresponding entropy. So the entropy
is just a function {\bf to} the event horizon.

When putting Eq. (\ref{3.12}) into Eq. (\ref{3.6}), we get the effective temperature of
higher-dimensional dS black holes in dRGT massive gravity,
\begin{equation}  \label{3.15}
T_{eff}=\frac{B(x, q)\left(1-x^{n+1}\right)}{4 \pi r_{c}
x^{n-1}\left(1+x^{n+2}\right)}.
\end{equation}
From Eq. (\ref{3.15}), we can draw the $T_{eff}-x$  curve of effective
temperature by using dimension $n=4,5,6$, certain $k$, and different $q$
taking the same or different $c_{0}^{2} c_{2} m^{2}$, $c_{0} c_{1} m^{2}$, $%
c_{0}^{3} c_{3} m^{2}$, $c_{0}^{4} c_{4} m^{2}$ when $r_{c}=1$, the
maximum of $T_{eff}$ is signed as $T_{eff}^{c}$, corresponding $x=x_{c}$; $T_{eff}=0$
corresponding $x_{min}=x_{0}^{T}$ as {\bf listed in T}able ~\ref{table1}.
\begin{table}[hbt]
\caption{When $n$ takes different values, the maximum values $T_{eff}$ under
different parameters are denoted as $T_{eff}^c$ corresponding $x=x_c$, and
the corresponding values $T_{eff}=0$ under different parameters are denoted
as corresponding $x=x_0^T$.}
\label{table1}\centering
{}
\begin{tabular}{|l|l|l|l|}
\hline
n & $\qquad x_{0}^{T} $ & $\qquad x_{c} $ & $\quad T_{eff}^c $ \\ \hline
4 & 0.0681925 & 0.0845323 & 4513.75 \\ \hline
5 & 0.12774 & 0.153701 & 3443.26 \\ \hline
6 & 0.186344 & 0.219437 & 3185.01 \\ \hline
\end{tabular}%
\end{table}

\begin{figure}[htp]
\centering
\subfigure[]{\includegraphics[width=0.225\textwidth]{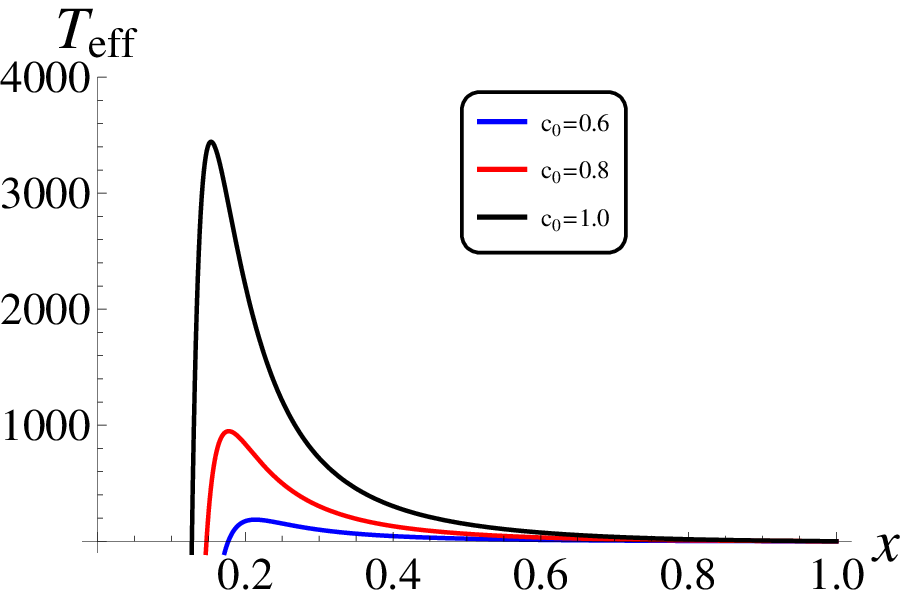}} %
\subfigure[]{\includegraphics[width=0.225\textwidth]{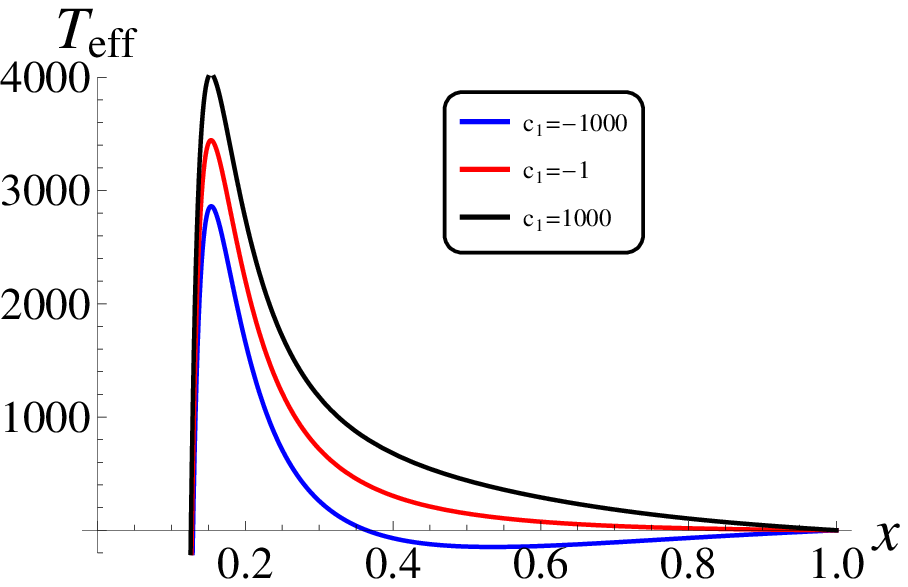}}\newline
\subfigure[]{\includegraphics[width=0.225\textwidth]{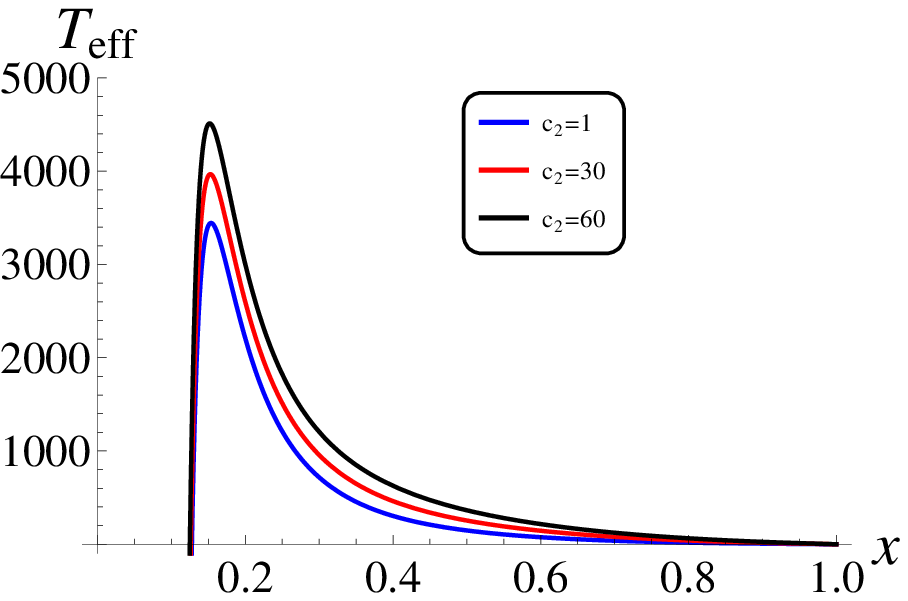}} %
\subfigure[]{\includegraphics[width=0.225\textwidth]{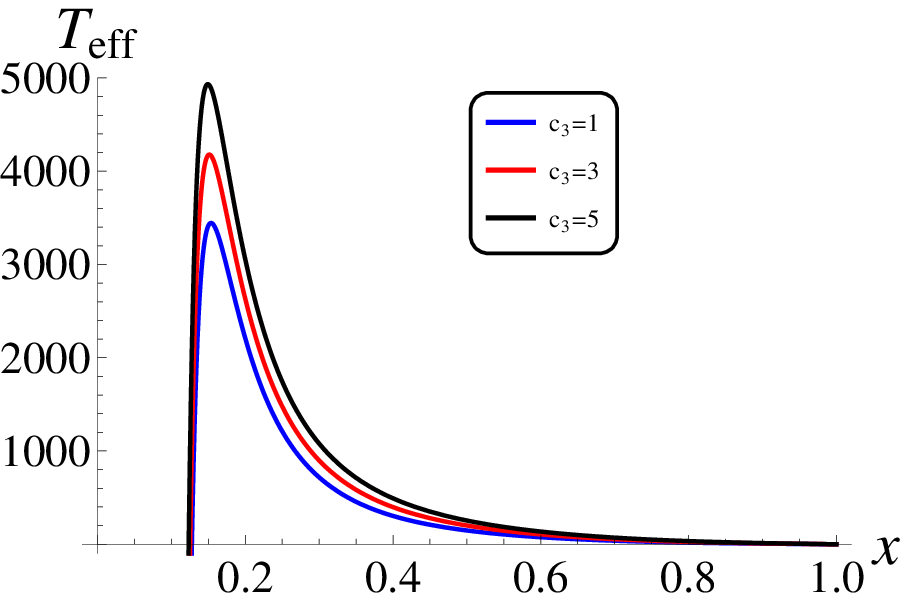}}\newline
\subfigure[]{\includegraphics[width=0.225\textwidth]{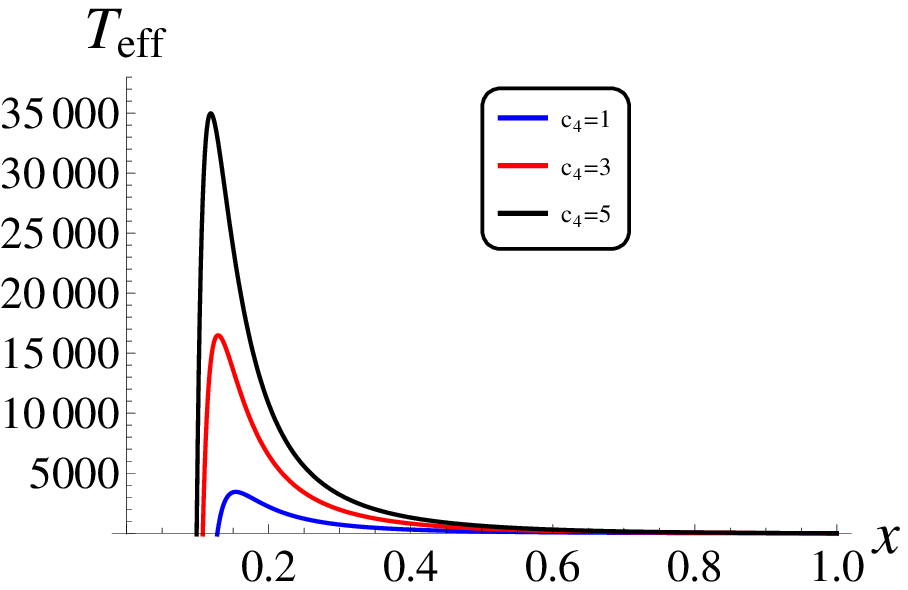}} %
\subfigure[]{\includegraphics[width=0.225\textwidth]{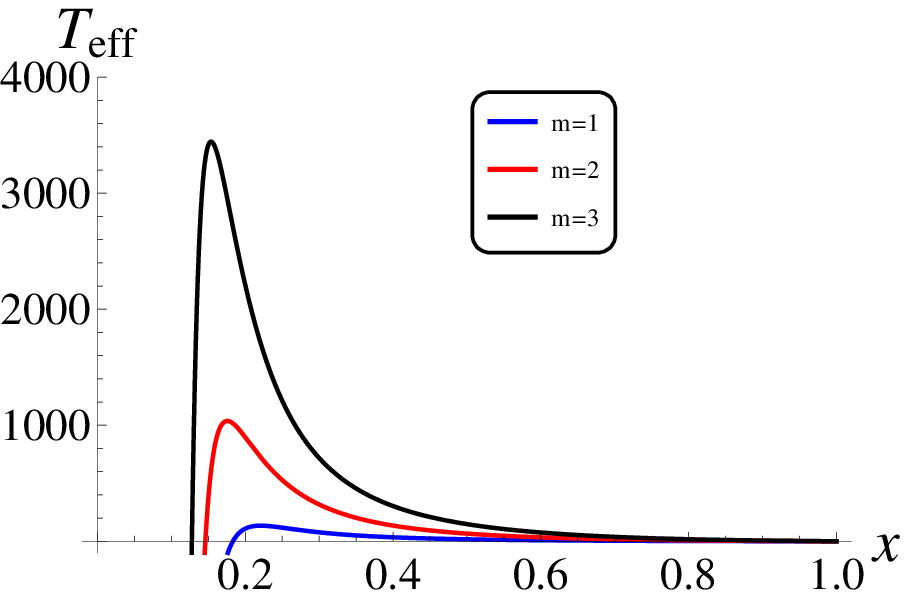}}\newline
\subfigure[]{\includegraphics[width=0.225\textwidth]{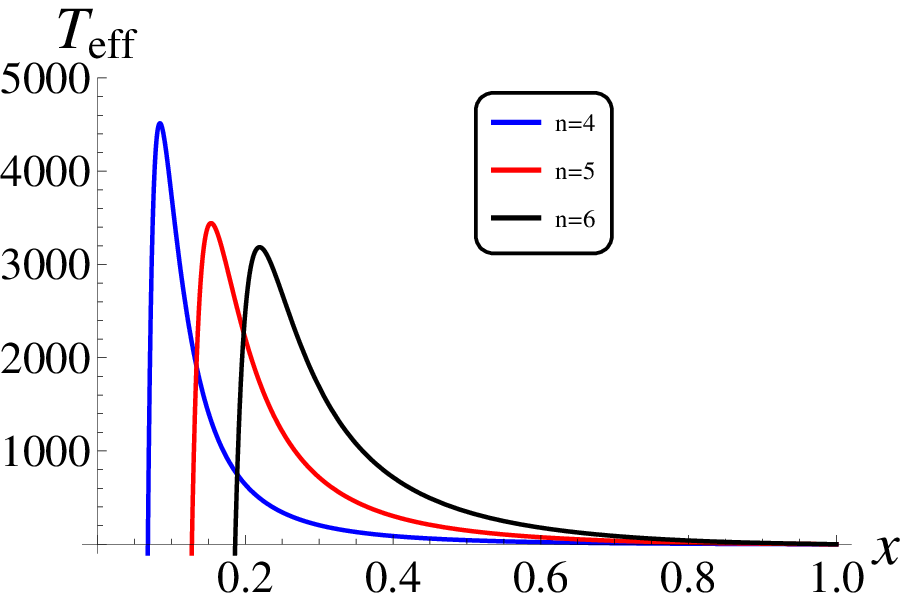}} %
\subfigure[]{\includegraphics[width=0.225\textwidth]{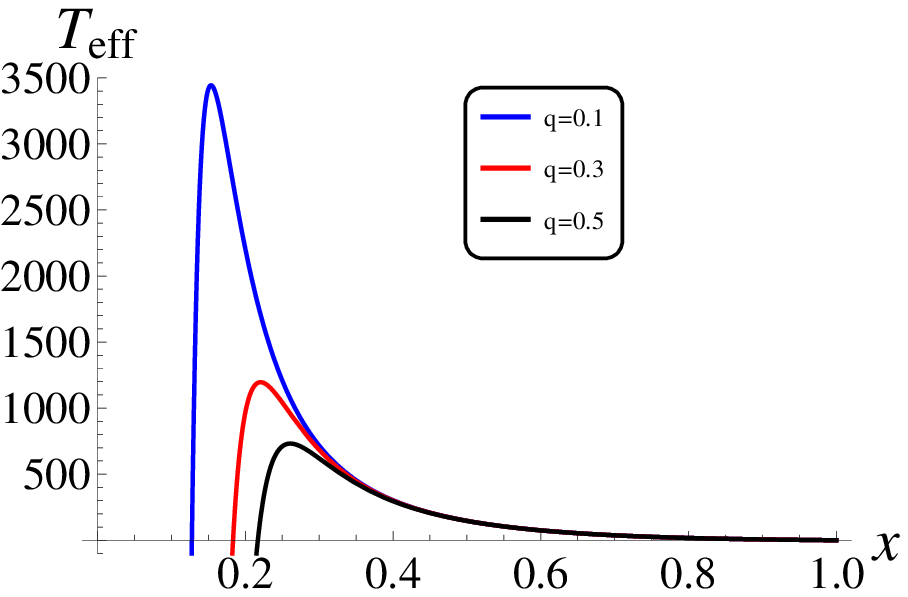}}
\caption{$T_{eff}-x $ diagrams when the parameters change respectively.}
\label{fig3}
\end{figure}

In order to clearly see the effect of relevant parameters on the effective
temperature, we illustrate an example of the $T_{eff}-x$ diagram with
different value of $c_0-c_4$ and $m$, $n$, $q$. which are explicitly shown
in Fig. \ref{fig3}(a)-\ref{fig3}(e). {\bf S}pecifically, the
maximum value of the effective temperature $T_{eff}$ of the system
increase{\bf s} with $c_0-c_4$, while the allowed region with $T_{eff}$ lager than
zero is also increased. Such tendency can also be seen from the behavior of
the effective temperature as a function of $x$, in term of different m{\bf ,} which {\bf is}
presented in Fig. \ref{fig3}(f). In Fig. \ref{fig3}(g), we can clearly see that the
maximum value of the effective temperature of the system will decrease with $%
n$, however the variable $x$ of maximum value of the effective temperature
is increasing. Meanwhile the whole interval with $T_{eff} > {\bf 0}$
moves to the right. More specifically, the maximum value of the effective
temperature of the system decrease{\bf s} with $q$, while the allowed region
with $T_{eff} > {\bf 0}$ is also reduced in Fig.  \ref{fig3}(h).

From Eq. (\ref{2.9}) Eq. (\ref{3.3}) and Eq. (\ref{3.5}), we get
\begin{equation}  \label{3.16}
\begin{aligned} P_{eff}&=\frac{n x^{n-1} g(x, q)}{2 k^{2} r_{c}^{2}
A(x)\left(1-x^{n+1}\right)^{2}}\\ &=\frac{g(x, q)}{2 k^{2}
r_{c}^{2}\left(1+x^{n+2}\right)\left(1-x^{n+1}\right)}, \end{aligned}
\end{equation}
where
\begin{equation}  \label{3.17}
\begin{aligned} g(x, q)=&\left[\left(k+c_{0}^{2} c_{2} m^{2}\right) \frac{n
\left(1-x^{2}\right)-\left(1-2 x^{n+1}+x^{2}\right)}{x}\right.\\
&+\frac{q^{2}\left[\left(1-x^{2 n}\right)-n\left(1+x^{2 n}-2
x^{n+1}\right)\right]}{2 n(n-1) r_{c}^{2 n-2} x^{2 n-1}}\\ &+\frac{c_{0}
c_{1} m^{2}}{n} r_{c}\left[n(1-x)-x\left(1-x^{n}\right)\right]\\ &+(n-1)
c_{0}^{3} c_{3} m^{2} \frac{n\left(1-x^{3}\right)-\left(2+x^{3}-3
x^{n+1}\right)}{r_{c} x^{2}}\\ &+(n-1)(n-2) c_{0}^{4} c_{4} m^{2} \\
&\quad\left.\frac{n\left(1-x^{4}\right)-\left(3+x^{4}-4
x^{n+1}\right)}{r_{c}^{2} x^{3}}\right] n F_{n}(x)\\ &-\left[\frac{n
x^{n-1}\left(1+x^{n+2}\right)}{\left(1-x^{n+1}\right)}-n x^{n}
F_{n}(x)\right]\\ &\quad\left[(n-1)\left(k+c_{0}^{2} c_{2}
m^{2}\right)\left(1-x^{2}\right)\right.\\ &\quad-\frac{q^{2}\left(1-x^{2
n}\right)}{2 n_{c}^{2 n-2} x^{2 n-2}}+c_{0} c_{1} m^{2} r_{c} x(1-x)\\
&\quad+(n-1)(n-2) c_{0}^{3} c_{3} m^{2} \frac{\left(1-x^{3}\right)}{r_{c}
x}\\ &\quad\left.+(n-1)(n-2)(n-3) c_{0}^{4} c_{4} m^{2}
\frac{\left(1-x^{4}\right)}{r_{c}^{2} x^{2}}\right]. \end{aligned}
\end{equation}

From Eq. (\ref{3.16}), we can draw the $P_{eff}-x$ curve parameters when
$r_{c}=1$, after being determined, and when $c_{0}^{2} c_{2} m^{2}$, $c_{0}
c_{1} m^{2}$, $c_{0}^{3} c_{3} m^{2}$, $c_{0}^{4} c_{4} m^{2}$, and any
parameter change in dimension and other parameters remain unchanged. We can
analyze the effect of parameters on the effective pressure $P_{eff}$. The
effect on the effective pressure with different $c_{0}^{2} c_{2} m^{2}$,
$c_{0} c_{1} m^{2}$, $c_{0}^{3} c_{3} m^{2}$, $c_{0}^{4} c_{4} m^{2}$.

\begin{figure}[htp]
\centering
\subfigure[]{\includegraphics[width=0.225\textwidth]{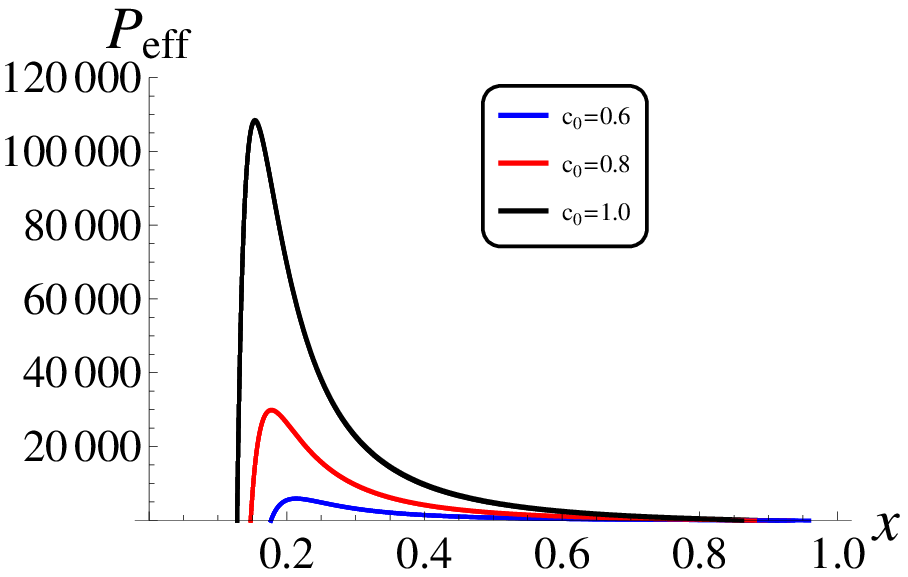}} %
\subfigure[]{\includegraphics[width=0.225\textwidth]{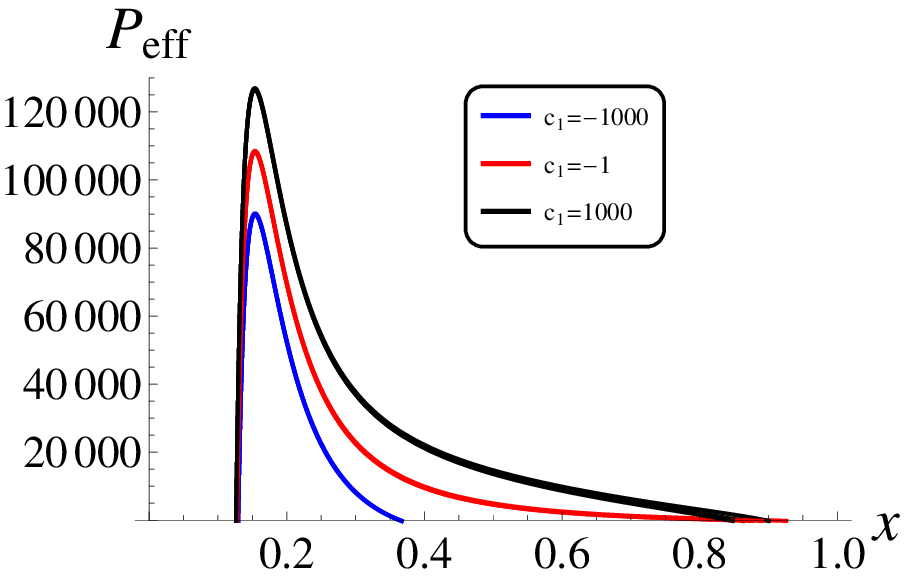}}\newline
\subfigure[]{\includegraphics[width=0.225\textwidth]{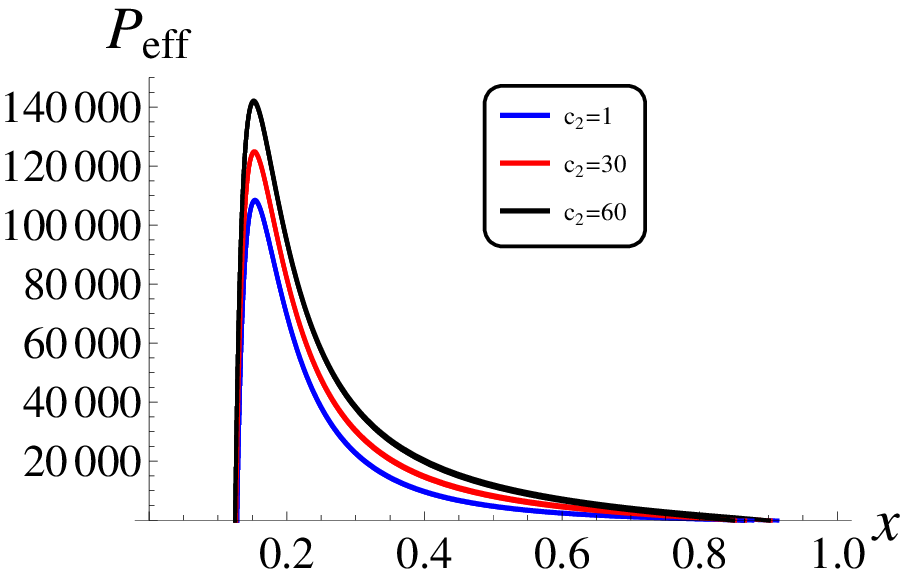}} %
\subfigure[]{\includegraphics[width=0.225\textwidth]{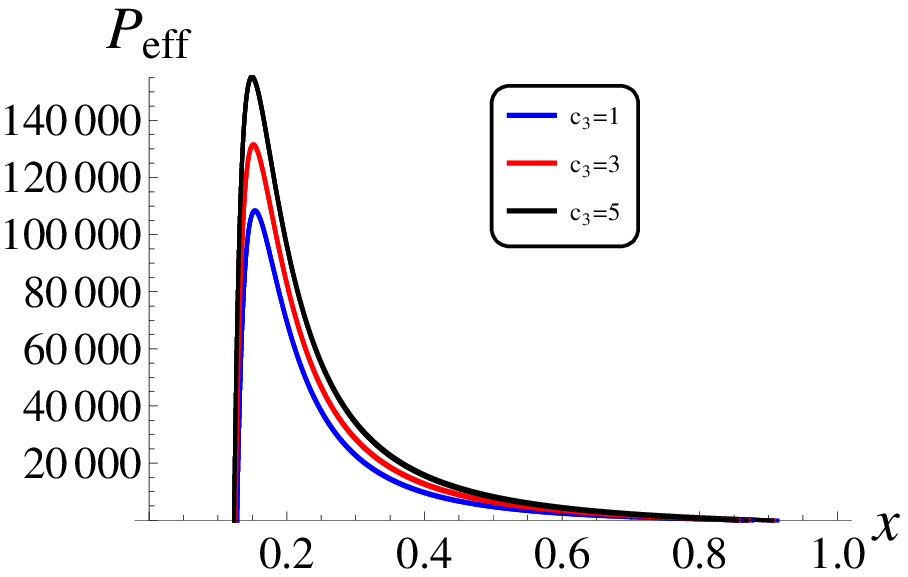}}\newline
\subfigure[]{\includegraphics[width=0.225\textwidth]{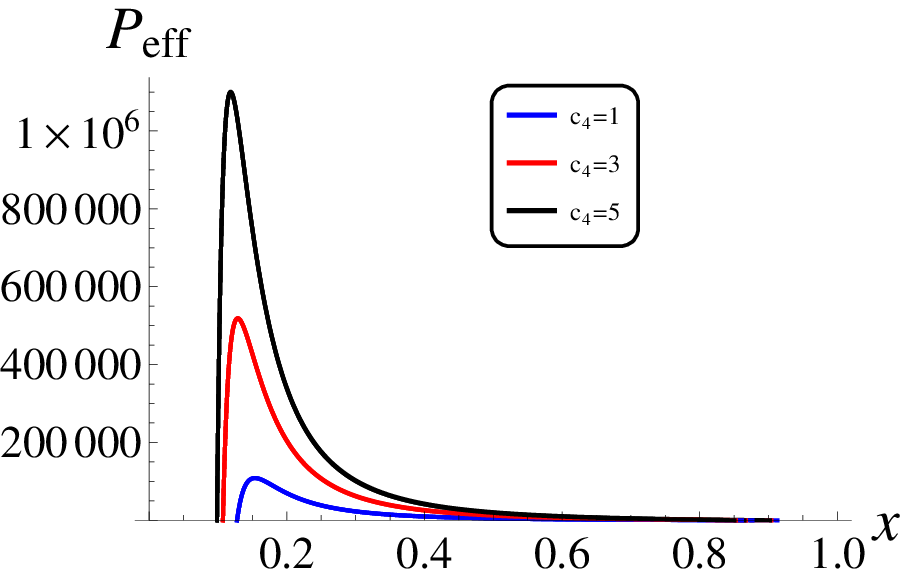}} %
\subfigure[]{\includegraphics[width=0.225\textwidth]{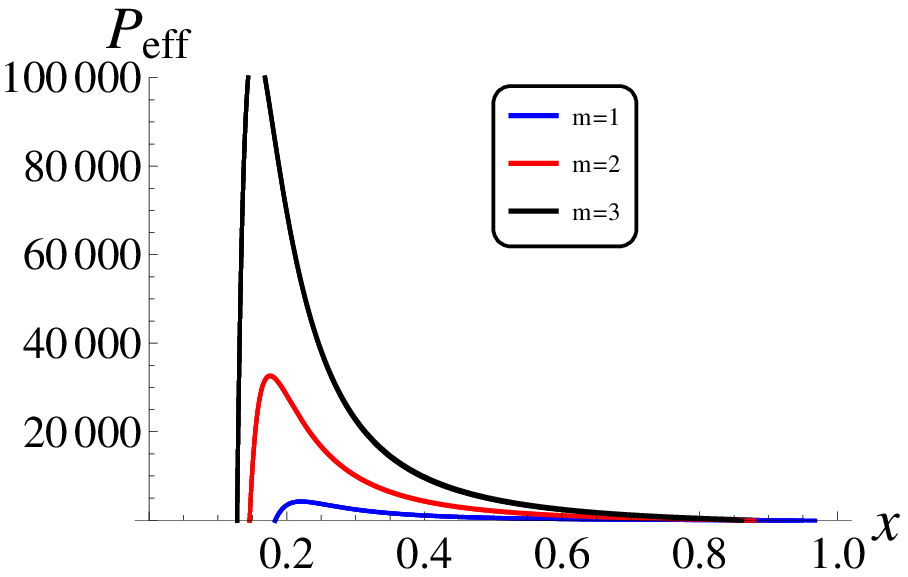}}\newline
\subfigure[]{\includegraphics[width=0.225\textwidth]{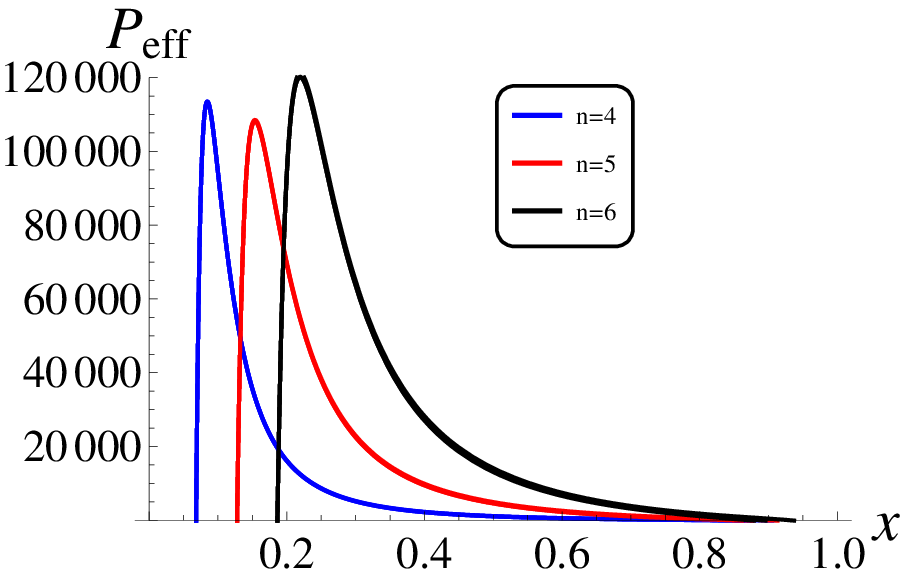}} %
\subfigure[]{\includegraphics[width=0.225\textwidth]{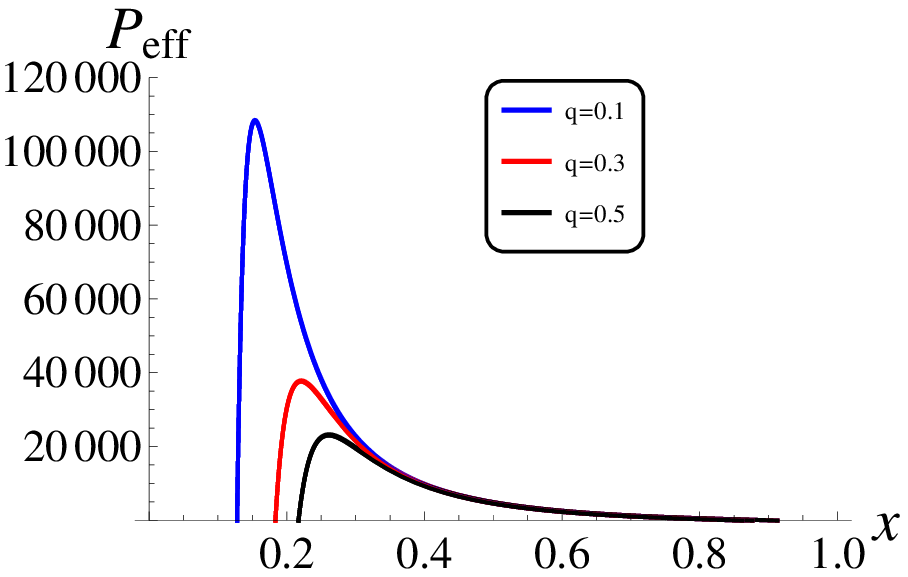}}
\caption{$P_{eff}-x$ diagrams when the parameters change respectively.}
\label{fig4}
\end{figure}

We also analyze the behavior of the effective pressure $P_{eff}$ with different value $c_0-c_4$, $m$, $n$, $q$ {\bf in Fig. \ref{fig4}} from which one could
clearly see the effect of these parameters on the effective pressure $P_{eff} $. Notice that it is the same behavior as that of the effective temperature shown in Fig. \ref{fig3}(a)- Fig. \ref{fig3}(e) and Fig. \ref{fig3}(h). Fig. \ref{fig4} shows the curve of the
effective pressure of the system changing with the parameters. Although the values of the curves vary with the parameters, the shapes of the curves are
{\bf very similar}. The curve has a maximum value $P_{eff}^c$ with the change of $x$, marked as $x_c$ and the effective pressure increases
monotonously {\bf as $x$ increasing} when $x_0^T<x<x_c$, while the effective pressure decreases monotonously {\bf as $x$ increasing in} the
range of $x_c<x<1$, the maximum value of the effective pressure $P_{eff}$ and its positive region ($P_{eff}>0$) increase with $c_0-c_4$ and $m$, {\bf but decreases with $q$}.
Differently, When $n$ is the
largest, $P_{eff}$ is also the largest, but $P_{eff}$ is not the smallest when $n$ is the smallest. When only $n$ increases, the $x$ corresponding to the largest $P_{eff}$ increases.


\section{Critical phenomena}

From Fig. \ref{fig3}, we can see that the curves of the effective
temperature of the system chang{\bf e} with the parameters. Although the values
of the curves vary with the parameters, the shapes of the curves are {\bf very similar}. The curve has a maximum value $T_{eff}^c$ {\bf as $x$ changes}, marked as $x_c$. The effective temperature increases
monotonously with the increase of $x$ when $x_0^T<x<x_c$ {\bf but} decreases monotonously with the increase $x$ at the
range of $x_c<x<1$. From the Fig. \ref{fig2}, the entropy
increases monotonously with $x$, so the heat capacity of the system is
positive in the $x_0^T<x<x_c$ interval, which satisfies the requirement of
equilibrium stability of the thermodynamic system, while the thermal
capacity of the interval system is negative at $x_c<x<1$ ranges, which does
not satisfy the requirement of qualitative thermodynamic equilibrium.

The expression of the heat capacity of a thermodynamic system
\begin{equation}  \label{4.1}
C=T_{eff}\left(\frac{\partial S}{\partial T_{eff}}\right),
\end{equation}
when $r_{c}=1 $, we put Eq. (\ref{3.3}) and Eq. (\ref{3.15}) into Eq. (\ref{4.1}), {\bf one} can
get curve $C-x$ .
\begin{figure}[htp]
\begin{minipage}[t]{0.45\textwidth}
   \centering
   \includegraphics[width=3in]{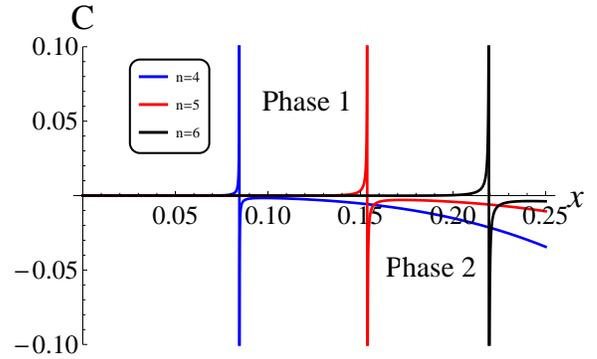}
    \caption{$ C-x $ diagram for $k=1, m=3, c_1=-3, c_2=1, c_3=1, c_4=1, r_c=1, q=0.1$.}\label{fig5}
\end{minipage}
\end{figure}

\begin{figure}[htp]
\begin{minipage}[t]{0.45\textwidth}
   \centering
   \includegraphics[width=3in]{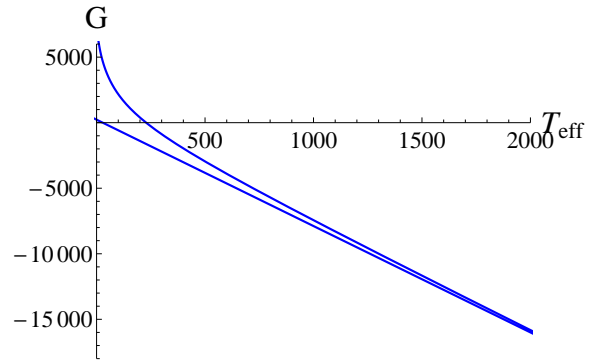}
    \caption{$ G-T_{eff}$ diagram for $n=5, k=1, m=3, c_1=-3, c_2=1, c_3=1, c_4=1, r_c=1, q=0.1$.}\label{fig6}
\end{minipage}
\end{figure}

In Fig. \ref{fig5}, Phase 1 is a steady-state interval, while phase 2 is a
non-steady-state interval. From the $C-x$ curve, the system diverges at the
point of $x=x_c$. From the Fig. \ref{fig2} and Eq. (\ref{3.2}), the
entropy and volume are continuous at the points $x=x_c$. According to
Ehrenfest's classification of phase transitions, the phase transitions
occurring in the system are second-order at the point $x=x_c$. In order to
further discuss the critical phenomena of the system, we discuss the Gibbs
free energy of the system, where $G=M-T_{eff}S$ \cite{Hendi17a,Zou17b}, and
draw the curves under isobaric conditions, as shown in Fig. \ref{fig6}.

From the Fig. \ref{fig6}, we know that {\bf the} effective temperature of the
system corresponds to two different Gibbs free energy values. According to
the irreversible process of the ordinary thermodynamic system under
isothermal and isobaric conditions, the Gibbs function always decreases.
Therefore, the actual process of the system follows the process of that the Gibbs function takes a small value. From the $C-x$ curve, in the increasing process of Gibbs' function, the heat capacity is negative, which means the system is thermodynamically unstable. {\bf S}o the black hole satisfying this $x_c<x<1$ condition is unstable, and there is no black hole satisfying th{\bf e} condition in the universe. When the effective temperature $T_{eff}$ of the black hole remains unchanged and the black hole is in the unsteady state region $x_c<x<1$ under external disturbance, the black hole
reaches the steady state region $x_0^T<x<x_c$ through the first-order phase
transition black hole. Therefore, the first-order phase transition occurring in de Sitter space-time is a process from unstable region to stable region,
which is different from the first-order phase transition of vdW system and
AdS black hole, because the first-order phase transition of vdW system and
AdS black hole are the transition between two states which meet the
requirements of thermodynamic equilibrium stability.

\section{Conclusion and discussion}

\label{sec:conclusion} Early studies on the effective temperature of de Sitter space-time were based on the assumption that the space-time entropy {\bf i}s known \citep{Urano09,Kanti17,Guo15,Guo16,Zhao14c,Ma15}, and the
effective temperature of de Sitter space-time was obtained by using the space-time thermodynamic quantity to satisfy the first law of thermodynamics. When the radiation temperature $T_{+}$ of the black hole
horizon is equal to that of the cosmological horizon: $T=T_{+}=T_{c}$, the effective temperature $T_{eff}$ obtained is not equal to that of the two horizons in general, that is $T_{eff}\neq T$, it is hard to be accepted. In addition, the hypothetical entropy in studying the effective temperature of de Sitter space-time has not been proved theoretically.

In this paper we obtain the entropy equation Eq. (\ref{3.12}) in HBHRGT space-time using the relationship of the thermodynamic first law, and the condition of $T=T_{eff}$ when black hole horizon radiation temperature is equal to cosmological space-time horizon's, that's $T=T_{+}=T_{c}$. In addition, we know that all parameters are not independent for multi-parameter space-time, because the radiation temperature of two horizons is equal, the space-time parameters need to satisfy equation Eq. (\ref{2.11}). However, when different independent parameters are taken, we all get the equation Eq. (\ref{3.12}) that the space-time entropy satisfies, which is independent of the parameters
selected, and it is the universal relationship between space-time. When $x\rightarrow 0$, space-time tends to be pure dS space-time (it only has the cosmological horizon), we get the HBHRGT space-time entropy function Eq. (\ref{3.13}) and the HBHRGT space-time effective temperature Eq. (\ref{3.15}) through solving differential equation. If we divide the HBHRGT space-time entropy $S$ into two horizons' sum $S_c+S_+$ and add interaction term $S_{t}$, we know that interaction is negative {\bf and positive value at $0<x<x_{0}$ and $x_{0}<x<1$, respectively,} from Fig. \ref{fig1} and the HBHRGT space-time entropy is {\bf an }increasing function of $x$, which is raising with $x$ as Fig. \ref{fig2}. The trend of curve changing is independent of space-time dimension, but the entropy is a function of dimensions of space-time. The curve of $S-x$, $T_{eff}$ and $P_{eff}$ are presented in Fig. \ref{fig2}, \ref{fig3} and \ref{fig4} respectively.

The curves also show the effect of the effective temperature $T_{eff}$ and {\bf the} pressure of each parameter $P_{eff}$. From Fig. \ref{fig3} of the curve, we can see that no matter how the parameters change, the effective temperature $T_{eff}$ of the system has a maximum $T_{eff}^{c}$, which is also the second-order phase transition temperature of the system. This characteristic is different from AdS black hole. From Fig. \ref{fig3} and \ref{fig4}, the
maximum value of the effective pressure $P_{eff}$ and the effective temperature $T_{eff}$, as well as its positive region ($T_{eff}>0, P_{eff} >0$) increase with $c_0-c_4$ and $m$ {\bf but} decrease with $q$. Differently, in Fig. \ref{fig3}(g), we can clearly see that the maximum
value of the effective temperature of the system decrease with $n$, however the variable $x$ of maximum value of the effective temperature is increasing. Meanwhile the whole interval with $T_{eff}$ lager than zero
moves to the right. While in Fig. \ref{fig4}(g), when only $n$ increases, the $x$ corresponding to the largest $P_{eff}$ increases.

The $C-x$ curve shows that the heat capacity has the positive value in the region $0<x<x_{c}$, which is satisfied with the requirements of thermodynamic system equilibrium stability. However {\bf it} is negative in the region $x_{c}<x<1$, {\bf which is} not satisfied the requirements of equilibrium stability. At $x=x_{c}$ point, the heat capacity is emanative, and {\bf the} system satisfies the requirements of the secondary phase change in thermodynamic system, so that $x=x_{c}$ is the second point of phase change. Therefore dS space-time is unstable in universe when $x$ is in range $x_{c}<x<1$. And there is only black hole that satisfies the points of $0<x<x_{c}$ possibly, which provides theoretical basis for {\bf one to investigate} black hole. From the $C-x$ curve, the effects of space-time dimension on the heat capacity and phase transition can be observed. When the parameters describing the space-time are fixed, the position of the phase transition point increases with the {\bf increasement} of the space-time dimension. Therefore the scope of the thermodynamically stable region increases correspondingly, which will lay the foundation for studying the thermodynamic properties of space-time {\bf in} higher dimension.

\section*{Acknowledgments}

We thank Prof. Z. H. Zhu for useful discussions.

This work was supported by the National Natural Science Foundation of China
(Grant Nos. 11847123, 11475108, 11705106, 11705107, 11605107), the Natural
Science Foundation of Shanxi Province, China (Grant No. 201601D102004). The
Initial Foundation of Mianyang Teachers' College (Grant No. QD 2016A002),
Natural Science Foundation of Education Department in Sichuan Province
(Grant No. 17ZB0210), the CQ CSTC under grant No.(cstc2018jcyjAX0192).

%

\end{document}